\documentclass[aps,pre,twocolumn,floats]{revtex4}
\usepackage{epsfig}
\usepackage{bm}

\begin{document}

\newcommand{\hide}[1]{}
\newcommand{\tbox}[1]{\mbox{\tiny #1}}
\newcommand{\half}{\mbox{\small $\frac{1}{2}$}}
\newcommand{\const}{\mbox{const}}
\newcommand{\ointt}{\int\!\!\!\!\int\!\!\!\!\!\circ\ }
\newcommand{\intt}{\int\!\!\!\!\int }
\newcommand{\ar}{\mathsf r}
\newcommand{\im}{\mbox{Im}}
\newcommand{\re}{\mbox{Re}}
\newcommand{\mbf}[1]{{\mathbf #1}}
\newcommand{\sinc}{\mbox{sinc}}
\newcommand{\trc}{\mbox{trace}}
\newcommand{\eexp}{\mbox{e}^}
\newcommand{\bra}{\left\langle}
\newcommand{\ket}{\right\rangle}


\title{Classical and quantum pumping in closed systems}

\author{Doron Cohen}

\affiliation{
Department of Physics, Ben-Gurion University, Beer-Sheva 84105, Israel
}

\date{Aug 2002, long published version \cite{pmp}, follow up \cite{pmo}} 



\begin{abstract}
Pumping of charge ($Q$) in a closed ring geometry is not
quantized even in the strict adiabatic limit. The deviation form
exact quantization can be related to the Thouless conductance.
We use the Kubo formalism as a starting point for the calculation
of both the dissipative and the adiabatic contributions to $Q$.
As an application we bring examples for classical dissipative
pumping, classical adiabatic pumping, and in particular we
make an explicit calculation for quantum pumping in case of the
simplest pumping device, which is a 3~site lattice model.
We make a connection with the popular $S$ matrix formalism
which has been used to calculate pumping in open systems.
\end{abstract}

\maketitle

Pumping of charge in mesoscopic \cite{marcus_rev}
and molecular size devices is regarded as a major
issue in the realization of future "quantum circuits"
or "quantum gates", possibly for the purpose of
"quantum computing".  Of particular interest is
the possibility to realize a pumping cycle that transfers
{\em exactly} one unit of charge per cycle
\cite{thouless,avronI,aleiner,barriers}.
In open systems this "quantization" holds only
approximately. But it has been argued \cite{aleiner}
that the deviation from quantization is due to
"dissipative" effect, and that exact quantization
would hold in the strict adiabatic limit, if the
system were {\em closed}.
In this Letter we would like to show that the
correct picture is quite different.
In particular we would like to make
a proper distinction between "dissipative" and
"adiabatic" contributions to the pumping, and
to calculate the deviation from exact quantization
in the latter case. As a starting point we adopt
the traditional Kubo formula \cite{landau},
but we also point out the relation
to the "adiabatic" \cite{berry,robbins}
and to the "$S$-matrix" \cite{BPT} formulations.
The present formulation of the pumping problem
has few advantages: It is not restricted  to the
adiabatic regime; It give a "level by level"
understanding of the pumping process;
It allows the consideration of any type of
occupation (not necessarily Fermi occupation);
It allows future incorporation of external
environmental influences such as that of noise;
It regards the "voltage" over the pump
as "electro motive force", rather than adopting
the conceptually complicated view \cite{ora}
of having a "chemical potential difference".
Finally, on the practical level, we give
a solution for the pumping
in a 3~site lattice model.
This is definitely the simplest pump circuit
possible, and we believe that it can be realized
as a molecular size device.
It also can be regarded as an approximation
for the closed geometry version of the
two delta potential pump \cite{barriers}.

The structure of this Letter is as follows:
We show how to get from the Kubo formalism
an expression for the pumped charge $Q$,
and explain the distinction between
"dissipative" and the "adiabatic" contributions.
Then we give illuminating examples
for classical dissipative pumping
and for classical adiabatic pumping.
Next we discuss the case of
quantum pumping, where the cycle is around
a chain of degeneracies. We show that
this can be understood as a special
case of "adiabatic transfer" scheme.
In order to get a quantitative estimate
for the pumped charge we consider
a 3~site lattice model. We get expressions
for $Q$, and express them in terms of the Thouless
conductance. We conclude by a short discussion
of the relation between the Kubo formalism,
the adiabatic formalism, and the $S$-matrix
formalism.


Consider a system that has a ring geometry (Fig.1a).
The Hamiltonian is ${\cal H}(x_1(t),x_2(t),x_3(t))$,
where $x_1$ and $x_2$ are parameters that control
the shape of the ring, or the height of some barriers,
while $x_3{=}\Phi{=}\hbar \phi$ is the magnetic flux.
We use units such that the elementary charge is unity.
The "generalized forces" are conventionally defined as
$F^k \equiv -{\partial {\cal H}}/{\partial x_k}$.
In particular $\langle F^{3} \rangle$ is the current $I$ through
the ring (see remark \cite{rmrk}).
Consider for a moment the time independent Hamiltonian
${\cal H}(x)$, with $x=\mbox{const}$,
and assume that the system is prepared in a
{\em stationary} state (either pure or mixed).
The expectation value $\langle F^k \rangle$
of a generalized force is known as the
"conservative force" or (in case of $k{=}3$)
as the "persistent current". The "fluctuations" of the
generalized forces are conventionally characterized
by the real functions:
\begin{eqnarray} \label{e1}
C^{ij}(\tau) &=& \langle \half (F^i(\tau)F^j(0)+F^j(0)F^i(\tau)) \rangle  \\
K^{ij}(\tau) &=& \frac{i}{\hbar} \langle [F^i(\tau),F^j(0)]\rangle
\end{eqnarray}
Note that both functions have a well defined classical limit.
Their Fourier transform will be denoted by $\tilde{C}^{ij}(\omega)$
and $\tilde{K}^{ij}(\omega)$ respectively.

Our interest in the following is in a {\em driving~cycle},
where $x=x(t)$ forms a loop in the 3~dimensional parameter space.
In linear response theory \cite{landau}
the non-conservative contribution to $\langle F^{k} \rangle$
is related to $x(t)$ by a causal
response kernel \mbox{$\alpha^{ij}(t-t')$}.
The Kubo expression for this response kernel is
$\alpha^{ij}(\tau) = \Theta(\tau) \ K^{ij}(\tau)$.
Its Fourier transform is the generalized
susceptibility $\chi^{ij}(\omega)$.
From here we can derive the expression
$\langle F^k \rangle =
-\sum_{j} \mbf{G}^{kj} \ \dot{x}_j$,
where $\mbf{G}^{kj}$ is the generalized
conductance matrix:
\begin{eqnarray} \label{e3}
\mbf{G}^{ij} \ = \ \lim_{\omega\rightarrow 0}
\frac{\im[\chi^{ij}(\omega)]}{\omega} \ = \
\int_0^{\infty} K^{ij}(\tau)\tau d\tau
\end{eqnarray}
Following Berry and Robbins \cite{robbins} we split
the conductance matrix
into symmetric and anti-symmetric parts.
Namely $\mbf{G}^{ij} = \bm{\eta}^{ij} + \mbf{B}^{ij}$.
The antisymmetric part $\mbf{B}$ can be regarded as a vector
$\vec{\mbf{B}}=(\mbf{B}^{23},\mbf{B}^{31},\mbf{B}^{12})$,
and the expression for the current can be written
in an abstract way as
$\langle F \rangle = -\bm{\eta} \cdot \dot{x} - \mbf{B}\wedge \dot{x}$.

The rate of dissipation, which is defined as the rate
in which energy is absorbed into the system, is given by
$\dot{{\cal W}} = -\langle F \rangle \cdot \dot{x} =
\sum_{kj} \bm{\eta}^{ij} \dot{x}_i\dot{x}_j$.
Only the symmetric part of $\mbf{G}^{ij}$
is responsible for dissipation of energy.
The adiabatic regime is defined by the condition
$|\dot{x}|  \ll  {\Delta^2}/{\hbar\sigma}$,
where $\Delta$ is the typical level spacing,
and $\sigma$ is the root mean square value of
the matrix element $(\partial{\cal H}/\partial x)_{nm}$
between neighboring levels.
In the adiabatic regime $\bm{\eta}^{ij}$ vanishes because
of the discreteness of the energy spectrum \cite{robbins}.
But outside of the adiabatic regime the levels
acquires an effective width
\mbox{$\Gamma/\Delta=(({\hbar\sigma}/{\Delta^2})V)^{2/3}>1$}
and therefore the smoothed version of $\tilde{K}^{ij}(\omega)$
should be considered. Consequently one can obtain
the fluctuation-dissipation (FD) relation:
$\bm{\eta}^{ij} \sim \tilde{C}^{ij}(\omega=0)$.
The formulation of the exact FD relation
depends on the assumptions regarding the
occupation $f(E_n)$ of the energy levels. See \cite{frc,wlf}.
Commonly one assumes a zero temperature Fermi
occupation, but this is not essential for the following analysis.
In order to derive the above expression for $\Gamma$
we have used the result of \cite{frc} (Sec.17) for the "core width"
at the breaktime $t=t_{\tbox{prt}}$ of perturbation theory.
Note that in the semiclassical limit (small $\hbar$) the
adiabaticity condition always breaks down.

The antisymmetric part $\mbf{B}$ of $\mbf{G}^{ij}$
does not have to vanish in the adiabatic limit. It can be
obtained from the adiabatic equation by looking
for a first-order stationary-like solution
\cite{thouless,avronI,robbins}, but we prefer to regard it
as a term in the (full) Kubo expression Eq.(\ref{e3}).
In \cite{berry,robbins} it has been demonstrated that it can be written as
\begin{eqnarray} \nonumber
\mbf{B}^{ij} &=& -2\hbar\sum_n f(E_n) \
\im\left.\left\langle\frac{\partial}{\partial x_i} n(x) \right|
\frac{\partial}{\partial x_j} n(x) \right\rangle
\\ \label{e4}
&=&
2\hbar \sum_{m \ne n} f(E_n)
\frac{\im\left[
\left(\frac{\partial {\cal H}}{\partial x_i}\right)_{nm}
\left(\frac{\partial {\cal H}}{\partial x_j}\right)_{mn}\right]}
{(E_m-E_n)^2}
\end{eqnarray}
Note that the "vertical" component of $\vec{\mbf{B}}$ vanishes
in the "horizontal" $x_3{=}0$ plane due to time reversal symmetry.

Disregarding a possible  persistent current contribution
(that does not exist in the case of a planar $\Phi{=}0$~cycle),
the expression for the pumped charge is:
\begin{eqnarray} \label{e6}
Q \ = \ \oint  I  dt \ = \
-\left[ \oint \bm{\eta} \cdot dx + \oint \mbf{B} \wedge dx \right]_{k=3}
\end{eqnarray}
If we neglect the first term, which is associated
with the dissipation effect, and average the second
("adiabatic") term over the flux, then we get
\begin{eqnarray} \label{e7}
\overline{Q|_{\tbox{adiabatic}}} \ = \
-\frac{1}{2\pi\hbar}\intt \mbf{B} \cdot \vec{dx} \wedge \vec{dx}
\ = \ \mbox{integer}
\end{eqnarray}
The integration should be taken over
a cylinder of vertical height $2\pi\hbar$,
and whose basis is determined by the projection of
the pumping cycle onto the $(x_1,x_2)$ plane.
The last equality is argued as follows:
The flux \mbox{$(1/\hbar)\intt \mbf{B}  \cdot  {dx}  \wedge  {dx}$}
through a surface that is enclosed by a cycle is the Berry phase \cite{berry}.
The result should be independent of the surface.
Therefore the flux through a {\em closed} surface
should equal $2\pi\times$integer.
Integrating over a cylinder, as in Eq.(\ref{e7}),
is effectively like integrating over a closed surface
(because of the $2\pi$~periodicity in the vertical
direction). This means that the flux averaged $Q$
of Eq.(\ref{e7}) has to be an integer.


Before we discuss the quantum mechanical pumping,
it is instructive to bring two simple examples for
{\em classical} pumping. In the following we consider
one particle  ($\mbf{r}$)
in a two dimensional ring as in Fig.1a.

The first example is for classical {\em dissipative} pumping.
The conductance $G=\mbf{G}^{33}$
can be calculated for this system \cite{wlf}
leading to a mesoscopic variation of the Drude formula.
The current is $I=-G\times\dot{\Phi}$, where $-\dot{\Phi}$
is the electro-motive-force.
Consider now the following pumping cycle:
Change the flux from $\Phi_1$ to $\Phi_2$,
hence pumping charge $Q = -G(1)\times (\Phi_2-\Phi_1)$.
Change the conductance from $G(1)$ to $G(2)$
by modifying the shape of the ring.
Change the flux from $\Phi_2$ back to $\Phi_1$,
hence pumping charge $Q(2) = -G(2) \times (\Phi_1-\Phi_2)$.
Consequently the net pumping
is $Q = (G(2)-G(1))\times(\Phi_2-\Phi_1)$.

The second example is for classical {\em adiabatic} pumping.
The idea is to trap the particle
inside the ring by a potential well
$U_{\tbox{trap}}(\mbf{r}_1{-}x_1(t),\mbf{r}_2{-}x_2(t))$.
Then make a translation of the trap along
a circle of radius $R$, namely
$x(t) = (R\cos(\Omega t),R\sin(\Omega t),\Phi{=}\const)$.
It is a-priori clear that in this
example the pumped charge per cycle is $Q=1$, irrespective
of $\Phi$. Therefore the $\vec{\mbf{B}}$ field must be
\begin{eqnarray} \label{e8}
\vec{\mbf{B}} = -\frac{(x_1,x_2,0)}{2\pi (x_1^2+x_2^2)}
\end{eqnarray}
This can be verified by calculation via Eq.(\ref{e4}).
The singularity along the $x_3$ axis
is not of quantum mechanical origin:
It is not due to degeneracies, but rather
due to the diverging current operator
($\partial {\cal H}/\partial x_3\propto 1/\sqrt{x_1^2+x_2^2}$).


We turn now to the quantum mechanical case.
Consider an adiabatic cycle that
involves a particular energy level $n$.
This level is assumed to have a degeneracy
point at $(x_1^{(0)},x_2^{(0)},\Phi^{(0)})$.
It follows that in fact there is
a vertical "chain" of degeneracy points
that are located at
\mbox{$(x_1^{(0)},x_2^{(0)},\Phi^{(0)}+2\pi\hbar\times\mbox{\small integer})$}.
These degeneracy points are important for the geometrical
understanding of the $\mbf{B}$ field, as implied by Eq.(\ref{e4}).
Every degeneracy point is like a monopole charge.
The total flux that emerges from each monopole
must be  $2\pi\hbar\times$integer for a reason that was
explained after Eq.(\ref{e7}).
Thus the monopoles are quantized in units of~$\hbar/2$.

The $\mbf{B}$ field which is created (so to say)
by a vertical chain of monopoles may have a different ``near field"
and ``far field" behavior, which we discuss below.
(Later we further explain that "near field"  means regions in $x$ space,
in the vicinity of degeneracy points, where $g_T \gg 1$,
while "far field" means regions where $g_T \ll 1$).
The far field regions exist if the chains are well isolated.
The far field region of a given chain is obtained by regarding the
chain as a smooth line.
This leads {\em qualitatively} to the same field as in Eq.(\ref{e8}).
Consequently, for a "large radius" pumping cycle
in the $\Phi=0$ plane, we get $|Q|\approx1$.
In the following we are interested in
the deviation from "exact" quantization:
If $\phi^{(0)}=0$ we expect to have
$|Q| \ge 1$, while if $\phi^{(0)}=\pi$ we expect $|Q|\le 1$.
Only for the $\phi$ averaged $Q$ of Eq.(\ref{e7})
we get {\em exact quantization}.

The deviation from $|Q|\approx 1$ is extremely large
if we consider a tight pumping cycle around
a $\phi^{(0)}=0$ degeneracy.
After linear transformation of the shape parameters,
the energy splitting $\Delta=E_n-E_m$ of the energy level~$n$
from its neighboring (nearly degenerated) level~$m$
can be written as
$\Delta=
((x_1{-}x_1^{(0)})^2+(x_2{-}x_2^{(0)})^2+
c^2(\phi{-}\phi^{(0)})^2)^{1/2}$
where $c$ is a constant. The monopole field is accordingly
\begin{eqnarray} \label{e9}
\vec{\mbf{B}} = \pm\frac{c}{2} \
\frac{( x_1{-}x_1^{(0)},  x_2{-}x_2^{(0)},   x_3{-}x_3^{(0)}) }
{((x_1{-}x_1^{(0)})^2+(x_2{-}x_2^{(0)})^2+
(\frac{c}{\hbar})^2(x_3{-}x_3^{(0)})^2)^{3/2}}
\end{eqnarray}
where the prefactor is determined by the requirement
of having a single ($\hbar/2$) monopole charge.
Assuming a pumping cycle of radius $R$ in the $\Phi=0$ plane
we get from the second term of Eq.(\ref{e6}) that
the pumped charge is $Q = \mp \pi \sqrt{g_T}$,
where $g_T=({\partial^2\Delta}/{\partial\phi^2})/\Delta={c^2}/{R^2}$
is a practical definition for the Thouless conductance
in this context. It is used here simply as a measure
for the sensitivity of an energy level to the magnetic flux $\Phi$.

What we want to do in the following is to "interpolate"
between the "near field" result, which is  $Q={\cal O}(\sqrt{g_T})$,
and the "far field"  result, which is $Q={\cal O}(1)$.
For this purpose it is convenient to consider
a particular model that can be solved exactly.
We consider a ring with two barriers.
The model is illustrated in Fig.2.
A version of this model, where the two barriers
are modeled as "delta functions", has been analyzed
in \cite{barriers} in case of {\em open} geometry.
Below we are going to analyze a different version
of the two barrier model, that allows an exact
solution for {\em closed} geometry.

We can classify the eigenstates of the {\em closed} ring
into two categories: wire states, and dot states (Fig.2a).
The latter are those states that are localized in the
"dot region" in the limit of infinitely
high barriers. In case of zero temperature Fermi
occupation we define $E_F$ as the energy of the last
occupied wire level in the limit of infinitely high barriers.
The two  "shape" parameters are the the bias $x_1$,
and the dot potential $x_2$.
The bias determines whether the dot tends
to exchange particles via the {\em left}
or via the {\em right} barrier.
The dot potential is loosely defined
as the energy of the dot level (Fig.2a).
A model specific definition of these
parameters in the context of the
3-site lattice Hamiltonian will be given later.

{\em The pumping cycle is assumed to be in the
$\Phi=0$ plane, so there is no issue
of "conservative" persistent current contribution}.
We start with a positive bias ($x_1>0$)
and lower the dot potential from a large $x_2>E_F$
value to a small $x_2<E_F$ value.
As a result, one electron is
transfered via the {\em left} barrier into the
dot region. Then we invert the bias
($x_1<0$) and raise back $x_2$. As a result
the electron is transfered back into
the wire via the {\em right} barrier.
A closer look at the above scenario (Fig.2b)
reveals the following:
As we lower the dot potential across a wire
level, an electron is adiabatically transfered
once from left to right and then from right
to left. As long as the bias is
positive ($x_1>0$) the net charge being pumped
is very small ($|Q| \ll 1$). Only the lowest wire
level that participate in the pumping cycle
carries $Q={\cal O}(1)$ net charge:
It takes an electron from the left side,
and after the bias reversal it emits it
into the right side.
Thus the pumping process in this model can be regarded
as a particular example \cite{avronI} of an {\em adiabatic transfer scheme}:
The electrons are adiabatically transfered from
state to state, one by one, as in "musical chair game".

For a single occupied level the net $Q$ is the
sum of charge transfer events that take place in
few avoided crossings. For many particle occupation
the total $Q$ is the sum over the net $Q$s
which are carried by individual levels.
For a dense zero temperature Fermi occupation
the summation over all the net $Q$s is a telescopic sum,
leaving non-canceling contributions only from the
first and the last adiabatic crossings. The latter
involve the last occupied level at the Fermi energy.

In order to get a quantitative estimate
for the $Q$ in a given avoided crossing,
we consider the simplest version
of the "two barrier model" that still
contains all the essential ingredients:
This is a three site lattice system.
The middle site supports a single "dot state",
while the two other sites support
two "wire states". The Hamiltonian is
\begin{eqnarray} \label{e10}
{\cal H} \mapsto \left(
\matrix{
0 & c_1 & \eexp{i\phi} \cr
c_1 & u & c_2 \cr
\eexp{-i\phi} & c_2 & 0} \right)
\end{eqnarray}
The three parameters are
the bias $x_1=c_1-c_2$,
the dot potential $x_2=u$,
and the flux $x_3 = \Phi = \hbar\phi$.
For presentation purpose we assume
that \mbox{$0 < c_1,c_2 \ll 1$}.
The eigenstates are $E_n$.
Disregarding the coupling between
the "wires" and the "dot" we have
two wire states with $E = \pm 1$,
and a dot state with $E = u$.
Taking into account the wire-dot
coupling we find that
there are two vertical chains
of degeneracies. The $u\approx-1$ chain is
\mbox{$(0,-1{+}c_1^2,2\pi\hbar\times\mbox{\small integer})$}
and the $u\approx1$ chain is
\mbox{$(0,+1{+}c_1^2,\pi+2\pi\hbar\times\mbox{\small integer})$}.

The eigenvalues $E_n$ are the solutions
of a cubic equation. Rather than
writing the (lengthy) analytical expressions
for them we give a numerical example for
their dependence on $u$ in the inset of Fig.3.
The eigenstates are
\begin{eqnarray} \label{e11}
|n(x)\rangle \mapsto
\frac{1}{\sqrt{S}}
\left(
\matrix{
c_2 \eexp{i\phi} + c_1 E_n \cr
1 - E_n^2 \cr
c_1\eexp{-i\phi} + c_2 E_n} \right)
\end{eqnarray}
where $S$ is the normalization.
Note that for $E=\pm 1$ we have
$S=2(c_1 \pm c_2)^2$, while
for $E=0$ we have $S\approx 1$.
After some algebra we find that
the first component of the
$\vec{\mbf{B}}$ field in the
$\Phi=0$ plane is
\begin{eqnarray} \nonumber
{\mbf{B}}^1 =
-2\im\left.\left\langle\frac{\partial}{\partial u} n(x) \right|
\frac{\partial}{\partial \phi} n(x) \right\rangle =
-(c_1^2-c_2^2)
\frac{1}{S^2}
\frac{\partial S}{\partial u}
\end{eqnarray}
Which is illustrated in Fig.3.
For a pumping cycle around
the \mbox{$u\approx \mp 1$} vertical "chain"
the main contribution to $Q$ comes from
crossing the $u\approx \mp 1$ line.
Hence we get
\begin{eqnarray} \label{e13}
Q = \pm \frac{c_1 \pm c_2}{c_1 \mp c_2} =
\pm\sqrt{1 \pm 2g_T}
\end{eqnarray}
where the Thouless conductance in this context
is defined as
$g_T = {2c_1c_2}/{(c_1 \mp c_2)^2}$.
In both cases we have approximate quantization
$Q=\pm 1 + {\cal O}(g_T)$ for $g_T \ll 1$,
while for a tight cycle either $Q \rightarrow \infty$
or $Q \rightarrow 0$ depending on which
line of degeneracies is being encircled.
If the pumping cycle encircles both "chains" then
we get $Q = {4c_1c_2}/{(c_1^2-c_2^2)}$.
In the latter case $Q={\cal O}(g_T)$ for $g_T \ll 1$,
with no indication for quantization.


For a pumping in a dot-wire system (see illustration in Fig.1b), 
in the limit of a very long wire (many sites) we express the Kubo 
formula for the conductance matrix using the $S$ matrix 
of the dot region. The derivation assumes ``quantum chaos", 
and leads to 
\begin{eqnarray} \label{e14}
\mbf{G}^{3j} = \frac{1}{2\pi i}
\trc\left(P\frac{\partial S}{\partial x_j}
S^{\dag}\right)
\end{eqnarray}
This is easily identified as the B\"{u}ttiker-Pr\'{e}tre-Thomas formula~\cite{BPT},
which has been derived for quantum pumping in open systems (Fig.1e).
In particular we get 
\mbox{$G^{33}=(1/(2\pi\hbar))\trc(PS(1{-}P)S^{\dag})$},
which is just the Landauer formula \cite{datta,fisher,stone}.

In summary we have shown how the Kubo formalism can
be used in order to derive both classical and quantum
mechanical results for the pumped charge $Q$ in
a closed system. In this formulation the distinction
between dissipative and non-dissipative contributions is manifest.
The dissipative contribution to the pumping
can be neglected in the adiabatic regime.
However, if the adiabaticity condition is violated
it does not mean automatically that we have
a dissipative effect. Classical pumping by translation
is an obvious example.
For the derivation of the
dissipative part of the Kubo formula it is essential
to realize that in generic circumstances
(unlike the case of translations)
the adiabatic equation does not possess
a stationary solution.


{\bf Acknowledgments:}
I thank Y. Avishai (Ben-Gurion University), Y. Avron (Technion)
and S. Fishman (Technion) for useful discussions.
This research was supported by the Israel Science Foundation (grant No.11/02).



\clearpage
\onecolumngrid

\centerline{\epsfig{figure=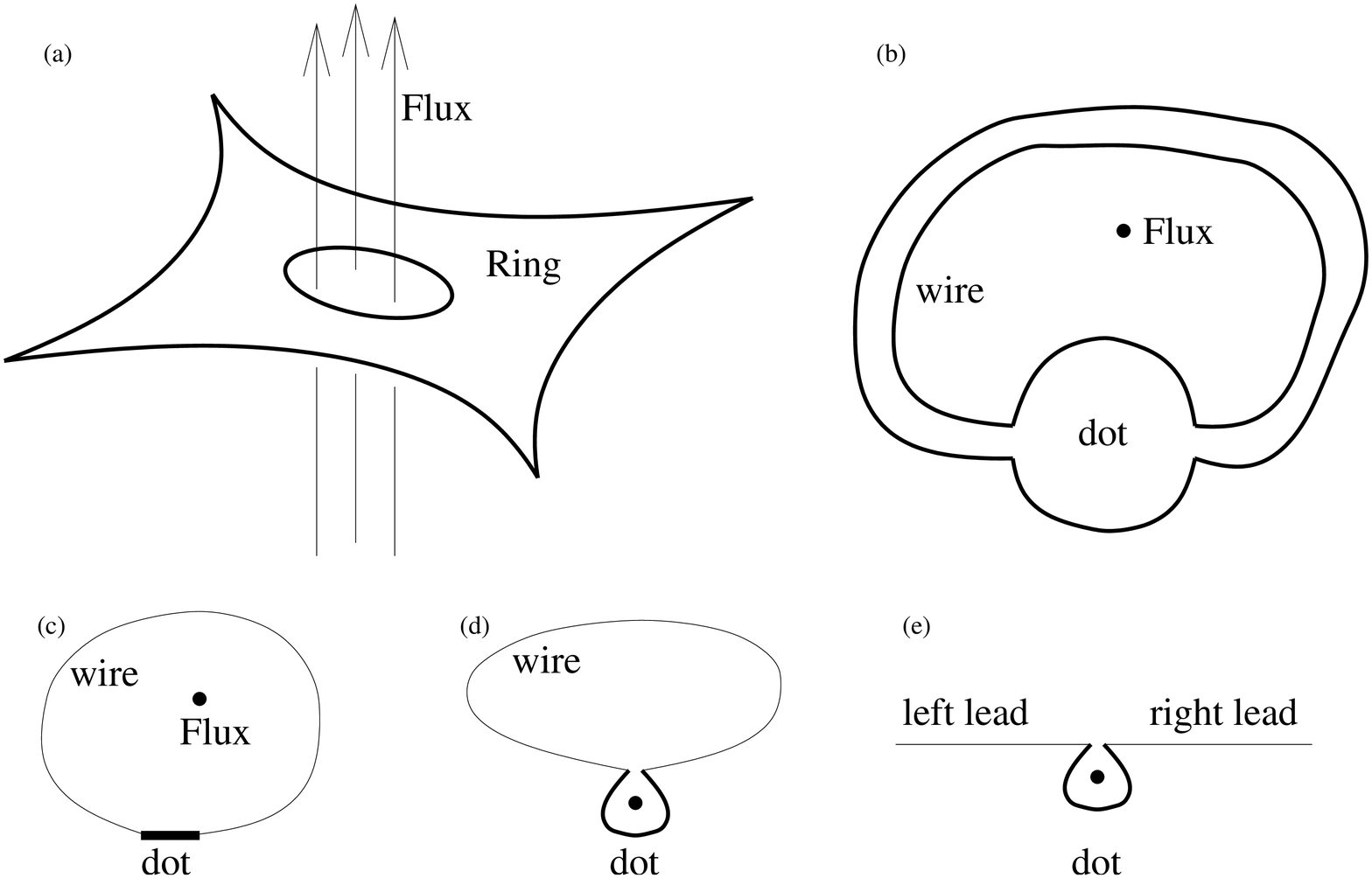,width=0.55\hsize}}
{\footnotesize FIG1.
Illustration of a ring system (a).
The shape of the ring is controlled
by some parameters $x_1$ and $x_2$.
The flux through the ring is $x_3=\Phi$.
A system with equivalent topology,
and abstraction of the model are
presented in (b) and (c).
The "dot" can be represented by an $S$ matrix
that depends on $x_1$ and $x_2$. In (d) also the
flux $x_3$ is regarded as a parameter of the dot.
If we "cut" the wire in (d) we get the open
two lead geometry of (e).}

\ \\ \ \\ \ \\

\centerline{
\epsfig{figure=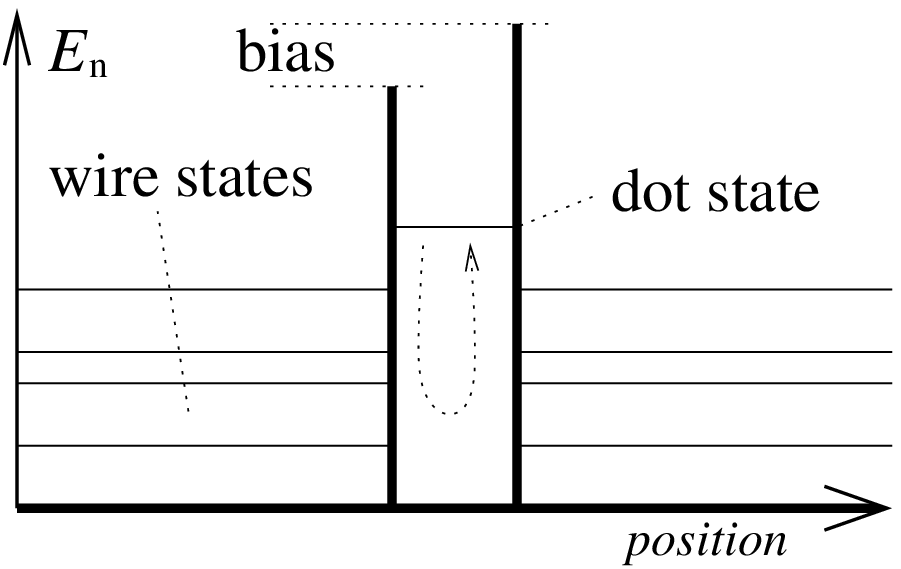,height=0.2\hsize}
\epsfig{figure=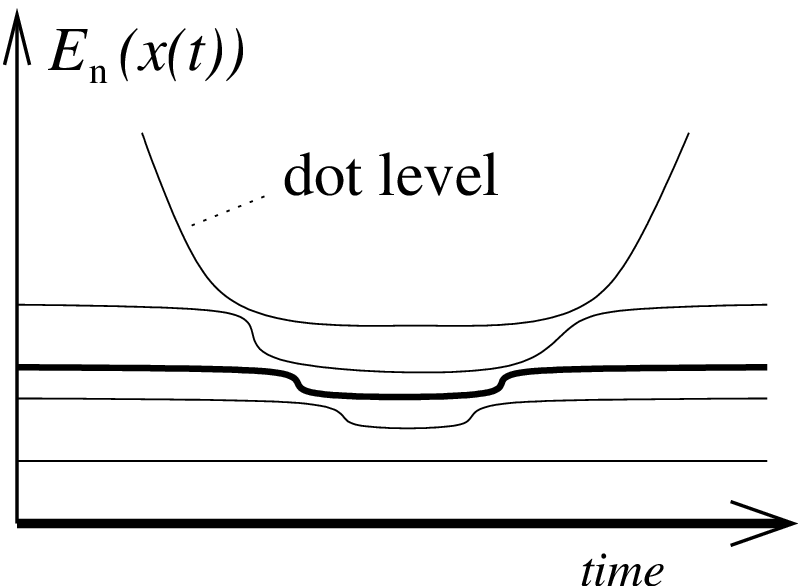,height=0.2\hsize}
}
{\footnotesize FIG2.
Schematic illustration of quantum pumping
in a closed wire-dot system. The net charge via the third
level (thick solid line on the right) is vanishingly
small: As the dot potential is lowered an electron
is taken from the left side (first avoided crossing),
and then emitted back to the left side
(second avoided crossing).
Assuming that the bias is inverted before the
dot potential is raised back, only the second level
carry a net charge $Q={\cal O}(1)$.}

\ \\ \ \\  

\centerline{\epsfig{figure=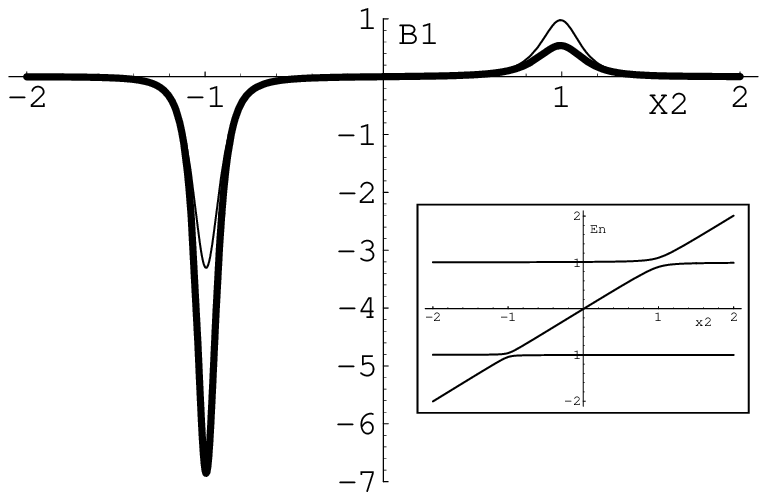,width=0.5\hsize}}
\
{\footnotesize FIG3.
The first component of the $\mbf{B}$ field
for a particle in the middle level of
the 3~site lattice model. It is plotted
as a function of the dot potential $x_2=u$.
The other parameters are $\phi=0$, and $c_1=0.1$,
while $c_2=0.04$ for the thick line and
$c_2=0.02$ for the thin line.
In the limit $c_2 \rightarrow 0$,
all the charge that is transfered from the
left side into the dot during the first avoided crossing,
is emitted back into the left side during the second
avoided crossing.
Inset: The eigenenergies $E_n(x)$ for
the $c_2=0.04$ calculation.}


\begin{thebibliography}{99}
\bibitem{rmrk}
There is some freedom in the
definition of the current operator.
For example one may define the current
through a given section. However,
the pumped charge of Eq.(\ref{e6}) comes
out the same due to the continuity equation.
\end{thebibliography}
\end{document}